# Instrumentation, acquisition and analysis of the Phase II SIMPLE dark matter search signals

**M. Felizardo**[1]
*Department of Physics, Faculty of Science and Technology,*
*New University of Lisbon, 2829-516 Caparica, Portugal*
*E-mail:* felizardo@itn.pt

**(for the SIMPLE Collaboration)**

I describe the new instrumentation for the SIMPLE dark matter search experiment, and its use in identifying, validating and rejecting non-WIMP backgrounds in the first stage of the Phase II project measurements. Beyond intrinsic acoustic background discrimination, evidence is provided for discrimination between $\alpha$- and neutron-induced events via analysis of the signal parameters. Analysis of the first stage result of the Phase II measurements yields 14 events associated with the ambient neutron field, consistent with MCNP simulations which include all materials radio-assays and full measurement shielding.

[1] Speaker

## 1. Introduction

The SIMPLE dark matter project [1,2] is based on Superheated Droplet Detectors (SDDs), a suspension of micrometric superheated droplets inside a viscous elastic gel, which undergo transitions to the gas phase upon energy deposition by incident radiation: each droplet behaves as a miniature bubble chamber. In Phase II of the project, the acoustic shock wave associated with the process of a bubble nucleation is recorded using new microphone-based instrumentation to detect and distinguish acoustic noise, gel-associated phenomena and particle-induced events. I describe the new instrumentation for the SIMPLE dark matter search experiment, and its use in identifying, validating and rejecting non-WIMP backgrounds in the first stage of the Phase II project measurements, comprising 15 superheated droplet detectors of total active mass 0.209 kg, with a total exposure, including losses from weather-induced power failures during the period, of 14.10±0.01 kgd.

## 2. Instrumentation and acquisition

The new instrumentation [3] is based on an omni directional high quality electret microphone cartridge (MCE-200) with a frequency range of 20 Hz – 16 kHz (3 dB), signal-to-noise ratio (SNR) of 58 dB and a sensitivity of 7.9 mV/Pa at 1 kHz. The microphone preamplifier (PGA 2500) is a digitally-controlled, analog preamplifier designed for front-end use with high performance audio analog-to-digital converters. The PGA 2500 features include low noise, a differential signal path and wide dynamic range. A factor 100 reduction in noise compared with the previous transducer instrumentation has been demonstrated.

Each detector contains feedthroughs for signal and pressure monitoring, and a microphone encased in a latex sheath, which is immersed in a 4 cm thick glycerin layer covering the gel at the top of the detector. The use of shielded telecommunications-grade cabling eliminates electromagnetic noise signals. Acoustic data is acquired in sequential Matlab files of ~ 6 MB each at constant rate of 8 kSps for periods of 8 minutes, and stored in one of two TB computers per 8 detector channels. Pressure data from each SDD is similarly acquired in a third TB computer.

## 3. Signal analysis

The signal records first undergo an event counting and validation routine: an amplitude threshold is set; the beginning and end of each spike is identified, based on the previous threshold; the time evolution of the spike is amplitude-demodulated; the decay time constant ($\tau$) of the pulse is measured; and finally, pulses which exhibit a $\tau$ below a given threshold are suppressed.

A discrimination filter analysis is next applied to all SDD signals: it establishes a threshold for each detector 2 mV above its noise level; rejects events that occur with the freon-less detector (acoustic noise); rejects all events with less than 5 pulse shape spikes; rejects acoustic duplicates on several detectors; rejects events out of the digital stable band-pass filter with the window of known nucleations (450-750 Hz); rejects events with $\tau$ which differ from the 5-40 ms of a true nucleation event; and finally identifies the surviving events though parameters such as Power Spectral Density (PSD), amplitude and time constants. A typical pulse shape of an event with its corresponding FFT can be seen in Fig. 1 and a comparison of true particle-induced nucleations with some acoustic backgrounds associated with the SDD gel dynamics in Table I [3].

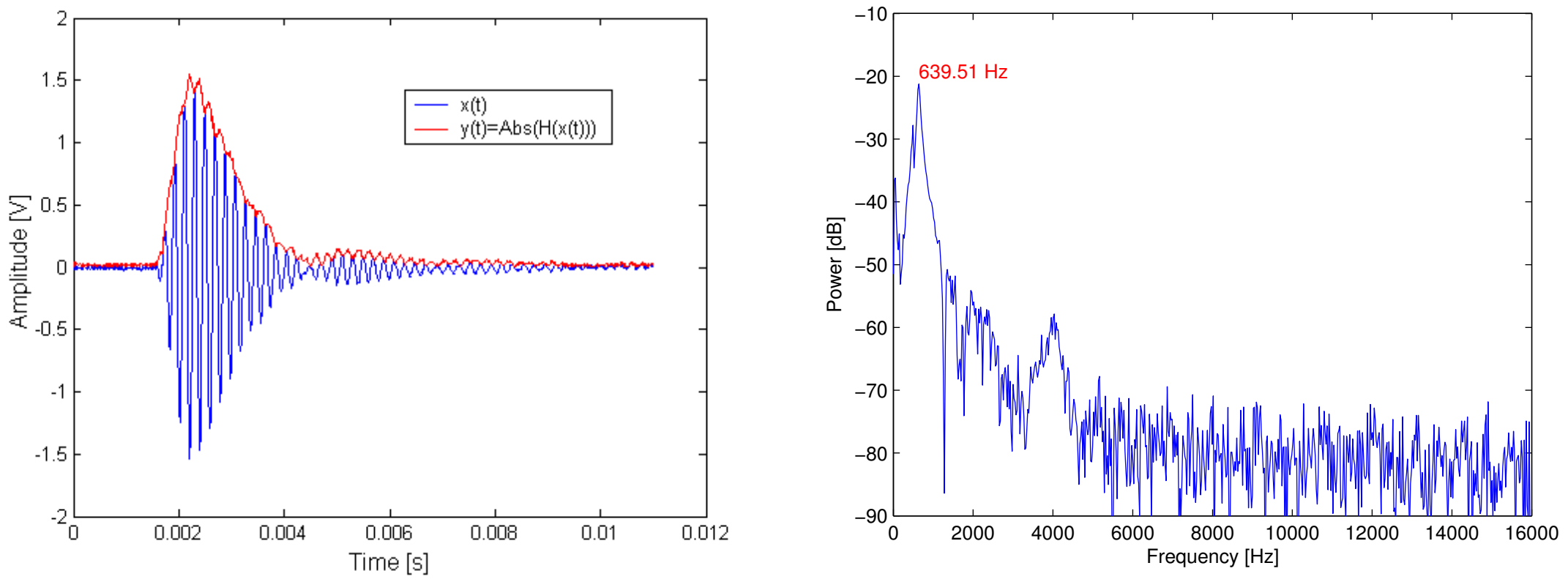

**Figure 1** (a) temporal evolution (pulse shape) of a typical particle-induced nucleation event; (b) FFT of the event shown in (a).

**Table I**: comparison of signal event characteristics with those of several common acoustic backgrounds.

| Event type | τ (ms) | frequency (Hz) | power (dB) |
|---|---|---|---|
| True nucleation | 5 – 40 | 450 – 750 | - 20±6 |
| Microleak | 10 -40 | 2800 – 3500 | - 25±10 |
| Fracture | 5 – 40 | 50 – 100 | - 25±8 |
| Trapped $N_2$ | 40 – 100 | 50 – 450 | - 55±3 |

The low noise and robustness of the preamp permits discrimination of nucleation events from acoustic backgrounds common to SDDs, including microleaks, fractures and trapped nitrogen gas in the gel of each detector.

## 4. Offline particle-induced calibrations

At 9°C and 2 bar, the reduced superheat of the SDDs is 0.3, and the probability of events from electrons, γ's and mip's was negligible over the exposure. Calibrations of the α response were made by doping the devices with $U_30_8$ during fabrication. The event signals, identified in the same fashion as described above, are shown in Fig. 2(a). Calibration of the SDD response to neutrons was performed using weak sources of Am/Be. The events are also displayed in Fig. 2(a), all of which occur with amplitudes ≤ 100 mV. Apart from providing a detection efficiency of 0.98±0.03, the neutron-induced events are seen to be of lower amplitude than the α-induced.

Figure 2(b) displays the corresponding frequency histogram of the squared calibration event amplitudes. The neutron population is well fit by a Gaussian plus constant background, from which a cut for A ≤ 100 mV gives an acceptance of > 97%.

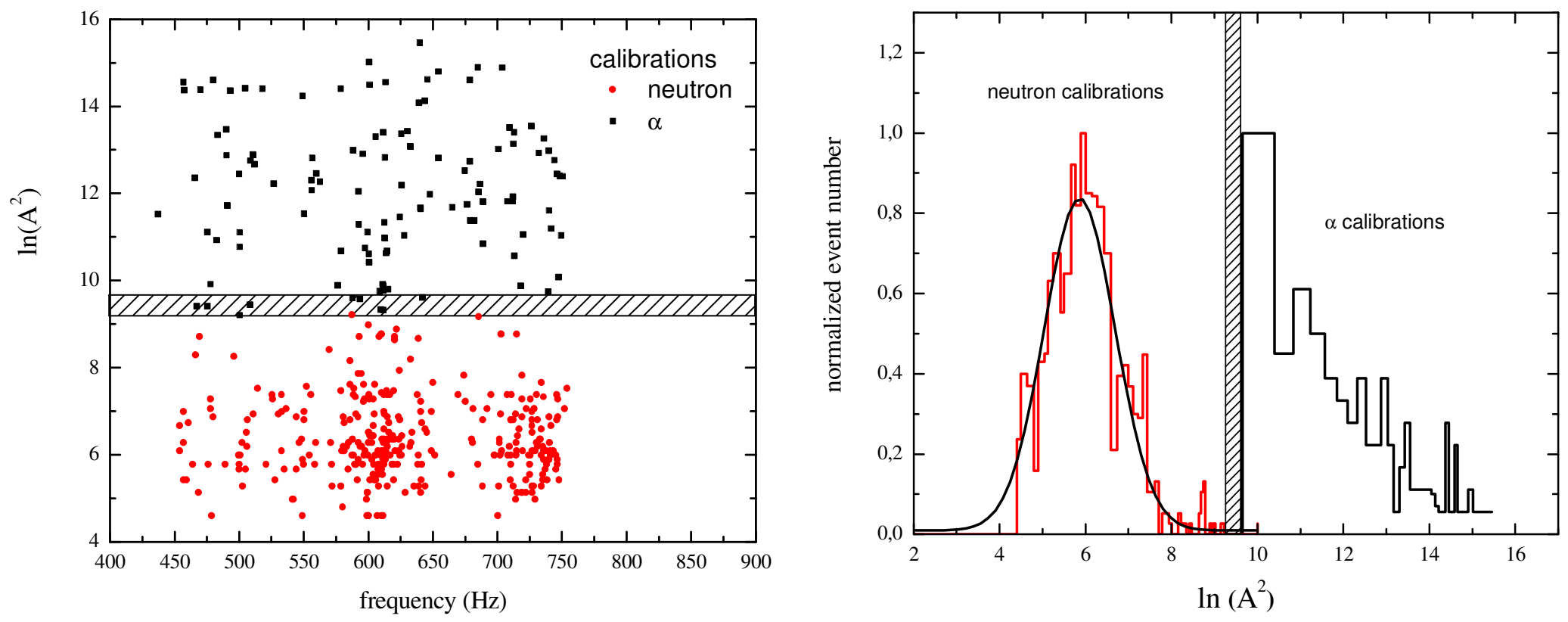

**Figure 2** (a) scatter plot of the $\ln(A^2)$ and frequency of each identified particle-induced calibration event; (b) frequency distribution of the calibration neutron and α amplitudes.

## 5. Results

The same analysis was applied to the Phase II run data, yielding 4058 events above the 2 mV detector threshold. Of these, 2228 were identified as coincidence events. The remaining events were examined on an event-by-event basis in terms of their signal and PSD characteristics, as shown in Fig. 3(a), isolating the true nucleation events with an efficiency of better than 97% at 95% C.L.

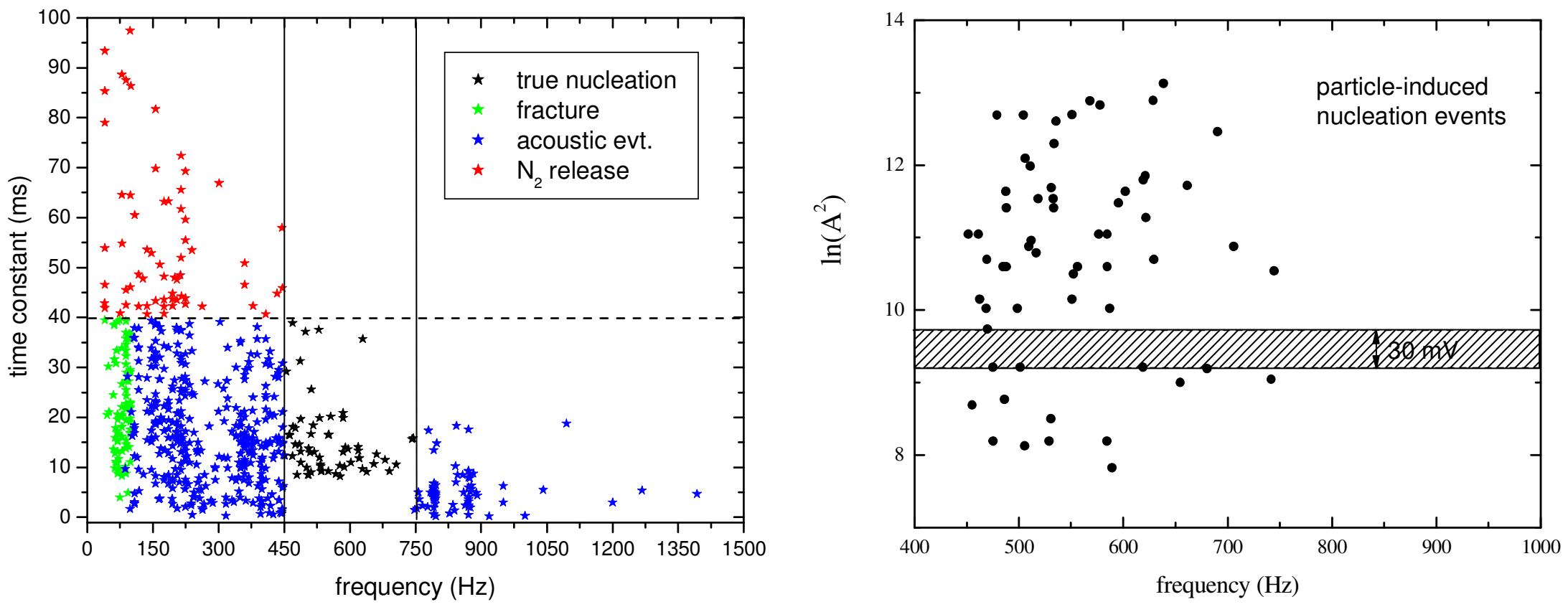

**Figure 3** (a) identification of the event origins from their power spectral density and time constants; (b) display of the $\ln(A^2)$ of each particle-induced event with the frequency of its primary harmonic (the hatched region indicates the 30 mV gap in the distribution); the difference with Fig. 2(b) results from the use of 15 cm water shield to enhance the tails of the moderated neutron spectrum.

Figure 3(b) displays the squared signal amplitudes (A) with frequency for each of the identified particle-induced events: a gap corresponding to A = 100-130 mV is discernible.

## 6. Conclusions

Efficient and low noise instrumentation based on a high-quality electret (MCE-200) microphone and its associated preamp (PGA-2500) has been implemented in the Phase II SIMPLE experiment. No significant variations in the instrumentation response were observed in testing at different temperatures and pressures. The low noise of the new instrumentation permits a clear discrimination of particle-induced events from acoustic backgrounds common to SDDs, including microleaks, fractures and trapped nitrogen gas.

The 46 events identified as $\alpha$-induced are consistent with estimates which include the measured radon field above the detectors, circulation of the tank water, diffusion of radon through the SDD materials, and $\alpha$-radio-assays of the detector construction materials. The 14 low amplitude events of the run identified from the neutron calibration measurements yield 0.99 ± 0.27 (stat) evt/kgd**, c**onsistent with MCNP simulations which include U/Th radioassays of the materials and the full experiment shielding [4], from which new restrictions on the WIMP-proton parameter space of spin-dependent WIMP interactions are obtained, as described elsewhere in these Proceedings [5].

## Acknowledgments

The work of M. Felizardo was supported by the Ph.D. scholarship program from the Fundação da Ciência e Tecnologia (SFRH/BD/46545/2008) of Portugal.

## References

[1] M. Felizardo*, et al.*, *First results of the Phase II SIMPLE dark matter search*, Phys. Rev. Lett. 105, 211301 (2010).

[2] T. Morlat, *et. al.*, *A CF3I-based SDD prototype for spin-independent dark matter searches*, Astrop. Phys. 30, 159 (2008).

[3] M. Felizardo*, et al.*, *New acoustic instrumentation for the SIMPLE superheated droplet detector*, Nucl. Instrum. & Meth. A589, 72 (2008).

[4] A.C. Fernandes, *et.al.*, *Studies on the efficiency of the neutron shielding for the SIMPLE dark matter search*, Nucl. Instrum. & Meth. A623, 960 (2010).

[5] TA Girard, *et al.*, *New limits on WIMP interactions from the SIMPLE dark matter search*, in proceedings of Identification of Dark Matter 2010 (IDM2010).